\documentclass[a4paper,twoside,twocolumn]{article}
\pdfoutput=1

\newif\iftr 
\trtrue     

\usepackage{ifpdf}  
\ifpdf
    \usepackage[pdftex]{graphicx}
    \usepackage[colorlinks,linkcolor=blue,citecolor=blue,urlcolor=blue]{hyperref}
    \DeclareGraphicsExtensions{.pdf}
\else
    \usepackage{graphicx}
    \usepackage[hypertex]{hyperref}  
\fi

\usepackage{todonotes} 

\usepackage{
url       %
,fancyhdr 
,lastpage 
,enumerate 
,amsmath  
,amsfonts 
,amssymb  
,multirow 
,xcolor   
,colortbl 
, enumitem 
}
\usepackage[hang]{footmisc} 
\usepackage{balance}

\usepackage[noindent, arraystretch, fullpage]{setlengths} 
\usepackage{
own         
}

\usepackage{amsthm}

\newtheorem*{test*}{Test}
\usepackage{xfrac}

\definecolor{Grey}{gray}{0.85}
\definecolor{good}{rgb}{0.85,1,0.85}
\definecolor{bad}{rgb}{1,0.85,0.85}
\newcolumntype{G}{>{\columncolor{Grey}}c}
\newcolumntype{g}{>{\columncolor{good}}c}
\newcolumntype{b}{>{\columncolor{bad}}c}

\graphicspath{{images/}}

\newcommand*{\metaauthori}{Bob Briscoe}
\newcommand*{\metashorttitle}{ECN Tunnelling Test}
\newcommand*{\metatitle}{A Test for IP-ECN Propagation by a Remote Tunnel Endpoint}
\newcommand*{\metano}{TR-BB-2023-003}
\newcommand*{\metakeywords}{Data Communication, Networks, Internet, Control, Active Queue Management, AQM, Pacing,  Medium Access Delay, MAC delay, Congestion Control, Quality of Service, Policing, Burstiness, Performance, Algorithm, Standards}

\newcommand*{\metamaili}{\href{mailto:research@bobbriscoe.net}{research@bobbriscoe.net}}
\newcommand*{\metaaddress}{}

\newcommand*{\metaversion}{01}
\newcommand*{\metadate}{28 Nov 2023}

\hypersetup{                       
     pdfauthor = {\metaauthori
     },
     pdftitle = {\metashorttitle},
     pdfsubject = {},
     pdfkeywords = {\metakeywords}
}%

\title{\metatitle}%
\author{\metaauthori%
\thanks{\metamaili, %
\metaaddress}%
\ %
}
\date{\metadate}%

\fancypagestyle{first}{%
\fancyhead[LO,RE]{}%
\fancyhead[LE,RO]{}%
\fancyhead[C]{}%
}%

\begin{document}
\bibliographystyle{alpha}%


\maketitle%
\thispagestyle{first}

\begin{abstract}
{\small\noindent%

This memo defines a brief set of tests to determine the decapsulation behaviour
of an unknown remote tunnel endpoint with respect to the Explicit Congestion
Notification (ECN) field in the Internet Protocol (IP) header. The tests could
be automated to be used by a tunnel ingress to determine whether the egress that
it is paired with will propagate ECN correctly.

}      
\end{abstract}
\section{Introduction}\label{ecn-encap-test_Intr}

This memo defines a brief set of tests to determine the decapsulation behaviour
of an unknown remote tunnel endpoint, with respect to the ECN field in the IP
header. It provides a table that says whether each possible detected behaviour
will propagate ECN correctly.

Test prerequisites are given in \S\,\ref{ecn-encap-test_Prereq}, the main hurdle
being the ability to overwrite the ECN field in the outer header at some point
along the span of a tunnel. This makes it hard to test `bump in the wire'
tunnels. To overcome this hurdle, a convenient arrangement would be to set up
the ingress of the tunnel under test on a host under the control of the tester.

The tests could be automated to be used by a tunnel ingress to determine whether
the egress it is paired with will propagate ECN correctly. Without such a test,
a tunnel ingress is required to zero the outer ECN field if it does not know
whether the egress it is paired with will propagate ECN
correctly~\cite{Briscoe16b:ecn-tunnel-scope_ID}.

In scenarios where there is no control protocol for a tunnel ingress to discover
the ECN capability of the egress, such a test could widen ECN coverage to
tunnelled paths where it is currently absent.

\subsection{Terminology}\label{ecn-encap-test_Term}

\begin{description}
	\item[Encap:] the encapsulation function at the ingress tunnel endpoint;
	\item[Decap:] The decapsulation function at the egress tunnel endpoint;
\end{description}

The following terms will be used for the IP header at different locations on the path relative to the tunnel, considering only the direction from application client to application server:
\begin{description}
	\item[Initial:] the header arriving at the tunnel ingress;
	\item[Inner:] the header that is encapsulated between the tunnel ingress and egress;
	\item[Outer:] the header that encapsulates the inner between the tunnel ingress and egress;
	\item[Onward:] the header leaving the tunnel egress.
\end{description}
\section{The Tests}\label{ecn-encap-test_Test}

\subsection{Test Prerequisites}\label{ecn-encap-test_Prereq}

\begin{itemize}[nosep]
	\item A working tunnel, e.g.\ a VPN;
	\item Access to one of the devices along the path of the tunnel, where the ECN
	field of the outer IP header can be altered\footnote{Ideally so it can be
	altered arbitrarily, but just being able to set congestion experiences (CE,
	i.e. 0b11) would support all the tests except one, which is a less important
	one anyway.};
	\item A remote application server, e.g.\ a web server (preferably a variety of
	different servers) that supports Accurate ECN feedback over either
	TCP~\cite{Briscoe14d:accecn_ID} or QUIC~\cite{Iyengar21:QUIC}.
	\item A local application client (e.g.\ a web browser), optionally with the
	ability to configure whether it sends ECN-capable packets (prior to
	tunnelling), and if so whether it sets ECT(0) or ECT(1).
\end{itemize}
\subsection{Test Setup}\label{ecn-encap-test_Setup}

Set up the tunnel as normal (procedure will depend on which type of tunnel).

\paragraph{If using TCP} configure the client TCP stack to use Accurate ECN (AccECN) feedback: 
\begin{description}[style=nextline, nosep]
	\item[Linux:] \texttt{\$ sysctl -w net.ipv4.tcp\_ecn=3}
	\item[MacOS:] \texttt{\$ sysctl -w net.inet.tcp.accurate\_ecn=1}
\end{description}

\paragraph{If using QUIC} make sure your QUIC implementation supports accurate
ECN feedback (at the time of writing, some still don't comply with the
spec~\cite{Iyengar21:QUIC}).

Make sure your application traffic is being routed via the tunnel.

\subsection{Control Test}\label{ecn-encap-test_Control}

The aim of this control test is to send packets with each of the four ECN
codepoints from the application client, then check that feedback from the
application server reflects the same codepoint.

Also it will be necessary to check that the tunnel ingress is copying each ECN
codepoint to the outer. If it's not, in order to test the remote tunnel
endpoint, it will be necessary to overwrite the outer with a copy of the Initial
ECN codepoint (using a similar approach to that for the main tests in
\S\,\ref{ecn-encap-test_Main}).

\paragraph{Details:} The Initial IP-ECN field can either be controlled by
configuring the client stack, or by overwriting the field in the packet before
it enters the tunnel. Given all codepoints cannot be set by configuration on all
packets, only the overwrite approach will be described here. One example
technique is to use the \texttt{tc} (traffic control) command to add a filter
that applies an action to packets matching the filter. The Linux \texttt{tc}
command is used for the example here, but  \texttt{tc} is also available for
MacOS.

\begin{verbatim}
$ tc filter add \
     dev DEV ingress flower MATCH_LIST \
     action pedit ex munge \
     ip dsfield set N retain 0x3
\end{verbatim}
where \texttt{N} would be respectively 0 to 3 to set the ECN field to Not-ECT,
ECT(1), ECT(0) or CE. \texttt{DEV} might be \texttt{eth0} for example. And an
example of a \texttt{MATCH\_LIST} might be \texttt{ip\_proto tcp dst\_port 80}
(see the tc-flower manual page for details).

To check the Outer (outgoing) ECN, and the server's (incoming) feedback of the
Onward ECN, Wireshark is recommended (version 4.0 onward supports AccECN in
TCP). For this control test, check that the Initial is the same as the feedback
of the Onward ECN, and that they are also the same as the Outer and the Inner.

The most specific feedback for testing purposes is given by TCP AccECN feedback
in the SYN-ACK from the server in response to the initial TCP SYN packet from
the client. The feedback is written with the `handshake encoding` into the three
ECN flags (AE, CWR, ECE) in the main TCP header as in the following table (from
Table 3 of \cite{Briscoe14d:accecn_ID}, which uses the TCP flags as newly
defined in Figure 2 of the same draft):

{\centering
\begin{tabular}{ccc}
	IP-ECN    & TCP-ECN  & Wireshark\\
	(outward) & (inward) &\\
	\hline%
	Not-ECT   & 0b010    & \texttt{.C.}\\
	ECT(1)    & 0b011    & \texttt{.CE}\\
	ECT(0)    & 0b100    & \texttt{A..}\\
	CE        & 0b110    & \texttt{AC.}\\
	\hline
\end{tabular}
\par}

If your client sends data packets to the server once the TCP connection has been
established, their feedback can be checked in AccECN TCP options that that
server sends to the client. These give a count of how many bytes of each
codepoint has been received by the server during the connection (counting from
1, not zero). However, they are not sent in response to every data packet (and
they are optional). So further explanation will not be given, but if the reader
wants to interpret this feedback, the definition of these TCP Options is in
\S\,3.2.3 of \cite{Briscoe14d:accecn_ID}.

QUIC feedback can also be checked, but it has to be decrypted first. Apple gives
instructions for how to allow Wireshark to decrypt QUIC for Cloudflare's quiche
stack in order to check the ECN 
feedback\footnote{\href{https://developer.apple.com/documentation/network/testing_and_debugging_l4s_in_your_app}{Testing and Debugging L4S in Your App}}, so that will not be repeated here. Then, any packet containing an ACK\_ECN frame can be viewed in Wireshark to read a count of the number of packets received by the server with each ECN codepoint: ECT(0), ECT(1) and ECN-CE.

\paragraph{Test robustness:} The test ought to be repeated a few times, and
preferably conducted with a few different application servers (but over the same
tunnel). This should help eliminate the possibility that:
\begin{itemize}[nosep]
	\item Active Queue Management (AQM) within the span of the tunnel is
	intermittently (and legitimately) setting the congestion experienced (CE)
	codepoint on the outer of some packets;
	\item A remote application server might have been chosen that provides
	incorrect ECN feedback due to an implementation bug.	
\end{itemize}

\begin{table*}
	{\centering
		\begin{tabular}{GGgggbb}
			\bf{Initial}&\bf{Outer}&\bf{RFC6040}&\bf{RFC4301}&\bf{RFC3168}&\bf{RFC2003}&\bf{mangled}\\
			&          &(unified)   &(IPsec)     &(original)  &(simple)    & \\
			\hline%
			Not-ECT     & CE       &dropped     &Not-ECT     & dropped    & Not-ECT    & \\
			ECT(1)      & CE       &CE          &CE          & CE         & ECT(1)     & \\
			ECT(0)      & CE       &CE          &CE          & CE         & ECT(0)     & \\
			ECT(0)      & ECT(1)   &ECT(1)      &ECT(0)      & ECT(0)     & ECT(0)     & \multirow{-4}{*}{other}\\
			\hline
		\end{tabular}
		\caption{Main Test: Possible Results and their Interpretation}\label{fig:Interpretation}
	}
\end{table*}

It should not be necessary to test both IPv4 \& IPv6 (and both combinations of
the two), because the definition of ECN is the same in both, so ECN processing
code should be common to both. However, a full test could include all four
combinations of IPv4 \& IPv6.

\subsection{Main Test}\label{ecn-encap-test_Main}

To test for correct operation of the remote tunnel egress, it is only necessary
to test the combinations in the first two (grey) columns of
\autoref{fig:Interpretation} (in addition to the control test above).
%

For this test, the filter action will need to be applied after tunnel
encapsulation. Then the outer will need to be overwritten with CE, for instance
using the \texttt{tc} command as already outlined in
\S\,\ref{ecn-encap-test_Control} with \(\mathtt{N}=\mathtt{3}\) (decimal), and
in the case of the last row, with \(\mathtt{N}=\mathtt{1}\).

\subsection{Interpretation of Results}\label{ecn-encap-test_Interpretation}

If the results conform with any of the green columns in
\autoref{fig:Interpretation}, the tunnel egress correctly propagates
ECN-marking, because it either complies with the latest ECN tunnelling spec
(RFC~6040~\cite{Briscoe07b:ECN-tunnel}) or with an earlier compatible spec
updated by RFC~6040 (IPsec~\cite{IETF_RFC4301:IPSEC_architecture} or the
original ECN spec~\cite{rfc3168}).

If, on the other hand, the results conform to one of the red columns, the tunnel
egress does not propagate ECN correctly. For instance, the first red column
shows the outcome of a `simple' tunnel, which just strips the outer on
decapsulation (as used before ECN tunnelling was first specified in 2001). The
final column `mangled' captures all other possible outcomes.

\newpage
\addcontentsline{toc}{section}{References}

{%
\scriptsize%
\bibliography{ecn-encap-test}}

\clearpage
\appendix
\section{History of ECN propagation by tunnels}\label{ecn-encap-test_History}

\begin{figure}[h]
	\centering
	\includegraphics[width=\linewidth,clip]{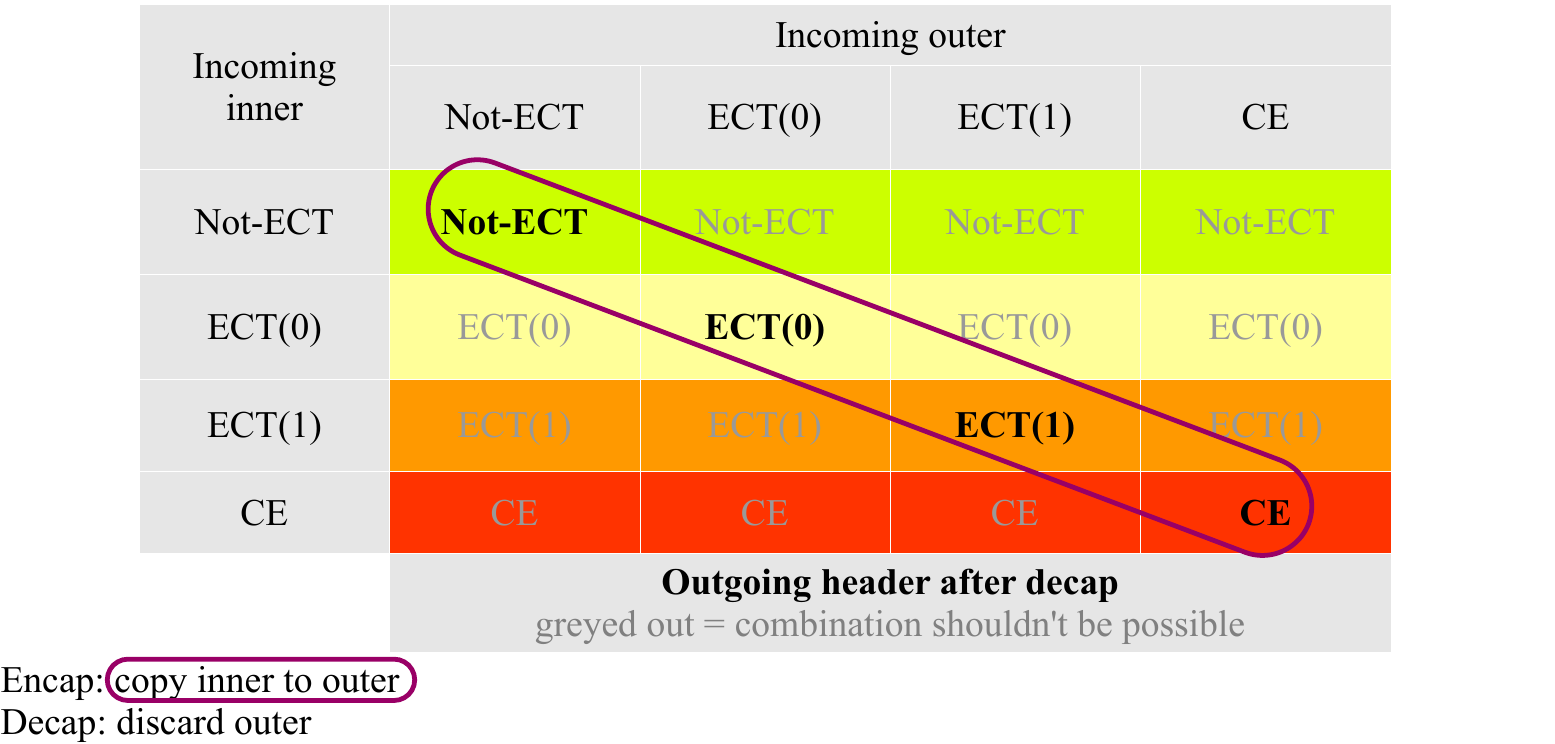}
	\caption{Simple (pre-ECN) tunnel, e.g.\ RFC2003}\label{fig:ecn-tunnel-testing-bg-1}
\end{figure}
\begin{figure}[h]
	\centering
	\includegraphics[width=\linewidth,clip]{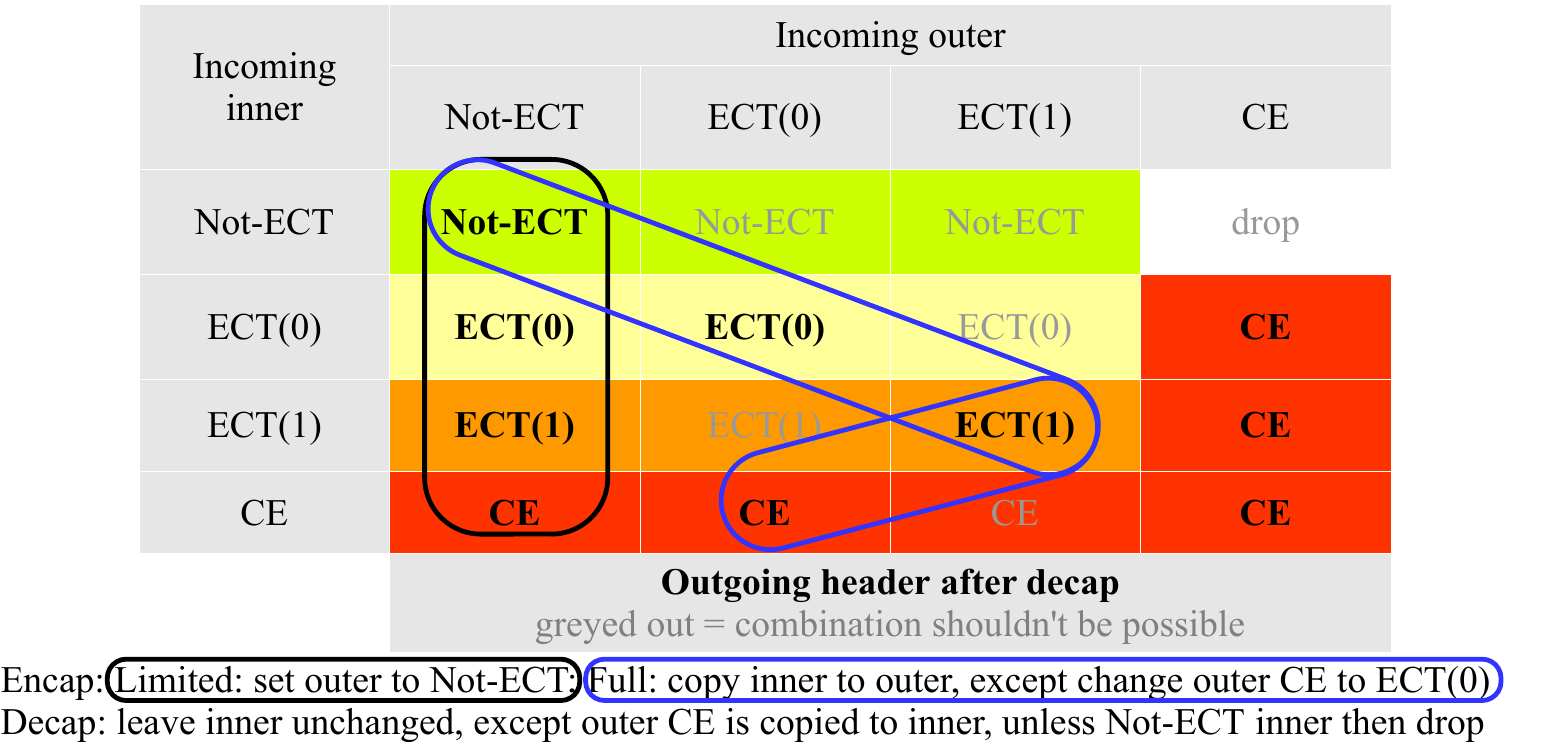}
	\caption{Original RFC~3168 ECN-tunnel~\cite{rfc3168}}\label{fig:ecn-tunnel-testing-bg-2}
\end{figure}

\begin{figure}[h]
	\centering
	\includegraphics[width=\linewidth,clip]{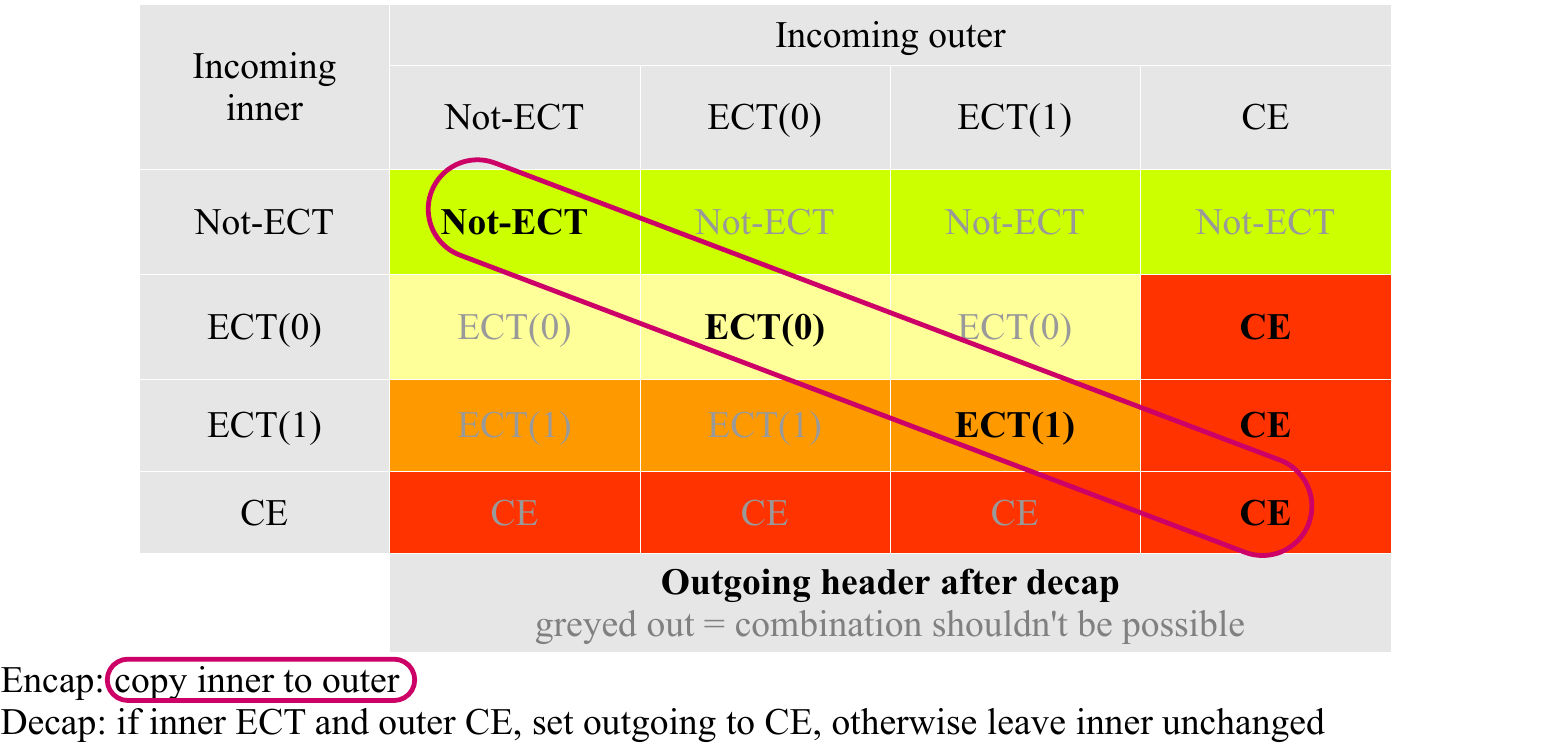}
	\caption{IPsec v2 (RFC~4301)~\cite{IETF_RFC4301:IPSEC_architecture}}\label{fig:ecn-tunnel-testing-bg-3}
\end{figure}

\begin{figure}[h]
	\centering
	\includegraphics[width=\linewidth,clip]{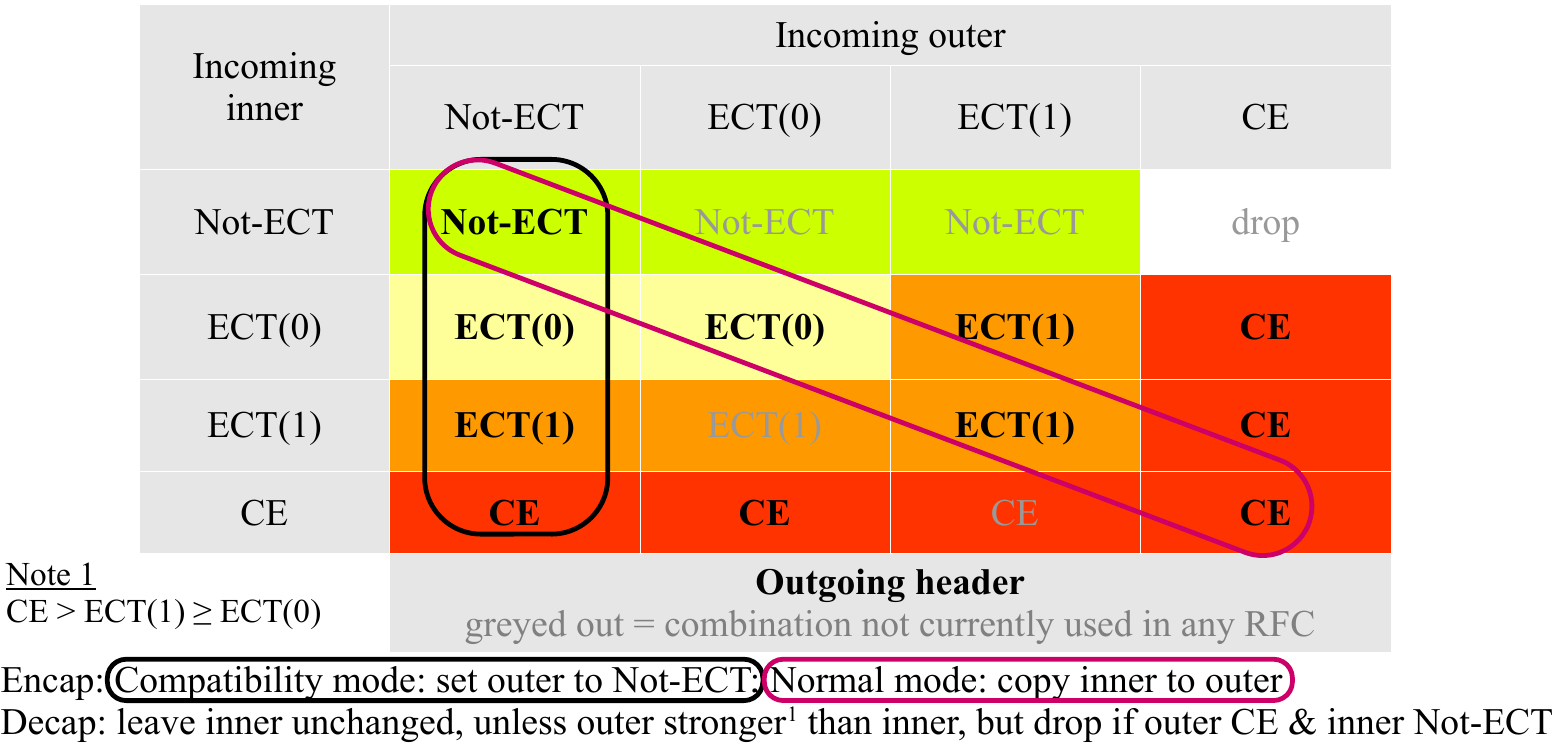}
	\caption{Universal ECN tunnel (RFC~6040)~\cite{Briscoe07b:ECN-tunnel}}\label{fig:ecn-tunnel-testing-bg-4}
\end{figure}

Figures \ref{fig:ecn-tunnel-testing-bg-1}--\ref{fig:ecn-tunnel-testing-bg-4}
illustrate the evolution of ECN tunnelling, starting from pre-ECN days in
\autoref{fig:ecn-tunnel-testing-bg-1}.

The table in each figure visualizes the outcome as each spec slightly altered
the decapsulation rules. The rows represent the Inner and the columns represent
the Outer header arriving at the tunnel egress. The text in each cell (and the
associated background colour) gives the Onward (outgoing) header.

The loops group together the combinations of Inner and Outer that would be
expected, given the behaviour of an encapsulator that complies with the same
spec as the decapsulator. Where the spec allows two encap options, different
coloured loops are shown for each.

\newpage

\begin{figure}[h]
	\centering
	\includegraphics[width=\linewidth,clip]{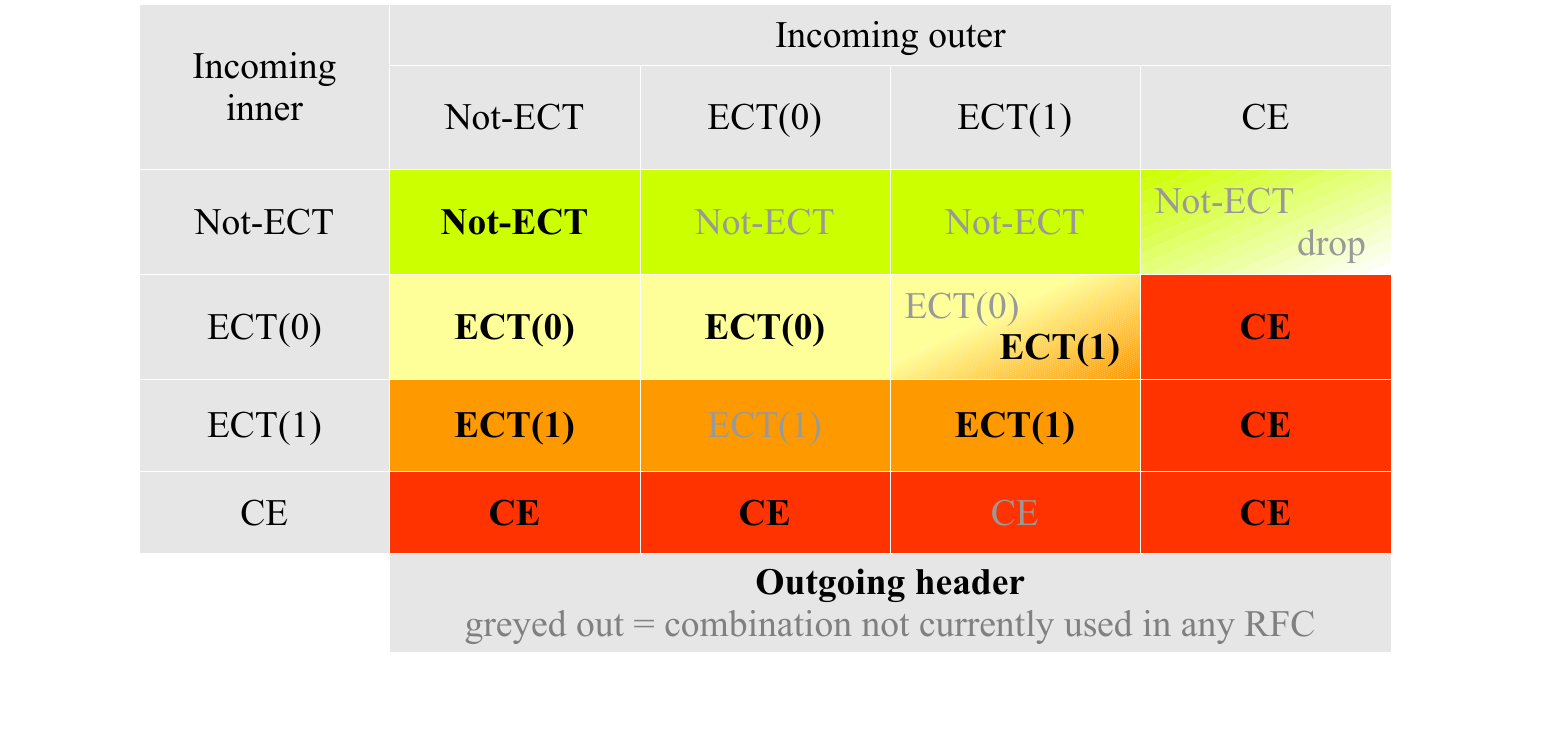}
	\caption{Black box view of all three combinations of ECN-tunnel specs (Figures \ref{fig:ecn-tunnel-testing-bg-2}--\ref{fig:ecn-tunnel-testing-bg-4})}\label{fig:ecn-tunnel-testing-bg-5}
\end{figure}

The text in cells outside the loops is greyed out to illustrate that this
combination would not be expected. Nonetheless, some other combinations of Inner
and Outer can occur when an encap complying with one spec is paired with a decap
complying with another. \autoref{fig:ecn-tunnel-testing-bg-5} overlays the three
behaviours that correctly propagate ECN to show how the three specs interact
with each other. It shows the union of all three possible encap behaviours as
not greyed out text, and two colours are used for the cell background where
there are two possible decap behaviours.

\onecolumn%

\onecolumn%
\addcontentsline{toc}{part}{Document history}
\section*{Document history}

\begin{tabular}{|c|c|c|p{3.5in}|}
 \hline
Version &Date &Author &Details of change \\
 \hline\hline
00A          &28 Nov 2023&Bob Briscoe &First Draft\\\hline%
\metaversion &\metadate  &Bob Briscoe &First Issue\\\hline%
\hline%
\end{tabular}

\end{document}